\documentclass[%
 preprint,
 superscriptaddress,
 amsmath,amssymb,
 aps,
 showkeys,
 titlepage,
 endfloats*.   
]{revtex4-2}

\usepackage[utf8]{inputenc}
\usepackage[english]{babel}

\usepackage{graphicx}
\usepackage{dcolumn}
\usepackage{bm}
\usepackage{booktabs}
\usepackage[colorlinks = true,
            linkcolor = blue,
            urlcolor  = blue,
            citecolor = red,
            anchorcolor = blue]{hyperref}

\begin{document}

\title[]{Unveiling the Reactivity of Oxygen and Ozone on C$_2$N Monolayer: A First-Principles Study}

\author{Soumendra Kumar Das}
\affiliation {School of Physical Sciences, National Institute of Science Education and Research (NISER) Bhubaneswar, HBNI, Jatni-752050, Odisha, India}
\author{Lokanath Patra}
\affiliation {Department of Mechanical Engineering, University of California Santa Barbara, CA, 93106, USA.}
\author{Prasanjit Samal}
\affiliation {School of Physical Sciences, National Institute of Science Education and Research (NISER) Bhubaneswar, HBNI, Jatni-752050, Odisha, India}
\author{Pratap K. Sahoo}
\email{psamal@niser.ac.in, pratap.sahoo@niser.ac.in}
\affiliation {School of Physical Sciences, National Institute of Science Education and Research (NISER) Bhubaneswar, HBNI, Jatni-752050, Odisha, India}
\affiliation {Centre for Interdisciplinary Sciences (CIS), NISER Bhubaneswar, HBNI, Jatni-752050, Odisha, India
}%

\date{\today}

\begin{abstract}
 The process of environmental oxidation is pivotal in determining the physical and chemical properties of two-dimensional (2D) materials. Its impact holds great significance for the practical application of these materials in nanoscale devices functioning under ambient conditions. This study delves into the influence of O$_2$ and O$_3$ exposure on the structural and electronic characteristics of the C$_2$N monolayer, focusing on the kinetics of adsorption and dissociation reactions. Employing first-principles density functional theory calculations alongside climbing image nudged elastic band calculations, we observe that the C$_2$N monolayer exhibits resistance to oxidation and ozonation, evidenced by energy barriers of 0.05 eV and 0.56 eV, respectively. These processes are accompanied by the formation of epoxide (C–O–C) groups. Furthermore, the dissociation mechanism involves charge transfers from the monolayer to the molecules. Notably, the dissociated configurations demonstrate higher bandgaps compared to the pristine C$_2$N monolayer, attributed to robust C-O hybridization. These findings suggest the robustness of C$_2$N monolayers against oxygen/ozone exposures, ensuring stability for devices incorporating these materials.
\end{abstract}

\maketitle

\section{\label{sec:level1}INTRODUCTION}
The emergence of two-dimensional (2D) materials with interesting and improved electronic, optical, plasmonic, thermoelectric, and mechanical properties compared to their bulk counterparts has created enormous interest in the research community in exploring novel physical phenomena for both fundamental physics and technological applications \cite{novoselov20162d,gupta2015recent,schaibley2016valleytronics,mannix2017synthesis,glavin2020emerging}. The introduction of quantum confinement due to reduced dimensionality in these systems has unlocked a wide array of applications, benefiting both industries and researchers~\cite{liu20192d,turunen2022quantum}. Among the large number of 2D materials studied so far including graphene, hexagonal boron nitride (h-BN), transition metal dichalcogenides, etc., nitrogenated holey graphene (C$_2$N) monolayer is a promising candidate having wide applications in optical devices \cite{guan2015effects, sun2016many}, catalysis \cite{li2016design, ma20163d}, molecular membranes \cite{zhu2015c}, photocatalysis \cite{mahmood2017efficient}, photovoltaics \cite{guan2017tunable}, photocatalytic water splitting \cite{Das2024}, hydrogen storage \cite{varunaa2019potential}, energy storage \cite{xu20172d}, gas separation \cite{yong2019c2n} etc. The presence of a direct band gap (1.96 eV), a porous structure with evenly distributed holes, high specific surface area, and well-aligned band edge positions spanning the water redox potential values makes the C$_2$N monolayer highly attractive.

The synthesis of two-dimensional C$_2$N structures through wet-chemical methods is well-established~\cite{mahmood2015nitrogenated}, with often introducing impurities like functional groups due to trapped small molecules (moisture, oxygen, etc.), leading to unexpected physical, chemical, and mechanical properties in resulting materials~\cite{wang2019elastic, patra2022surface}. Interestingly, the 2D C$_2$N crystal exhibits semi-metallic (graphene-like) behaviour before annealing, attributed to unintentional doping effects caused by trapped impurities and/or adsorbed gases within the porous crystal structure, suggesting tunable electronic properties~\cite{mahmood2015nitrogenated}. Additionally, the comprehension of oxygen (O$_2$) and ozone (O$_3$) adsorption and dissociation on new materials is also crucial from health and environmental perspectives, along with their impact on diverse devices, considering their extensive utilization in the semiconductor industry and various sectors such as food, water treatment, textiles, pharmaceuticals, and medical fields ~\cite{mcdonnell2019uv, wang2017physics, vig1976uv, vig1985uv, jourshabani2022efficient, de2000detailed, sankaranarayanan2010electric, perry2011decontamination}. Furthermore, the catalytic activity of 2D monolayers is significantly influenced by the adsorption and dissociation of O$_2$ molecules. Hence, understanding the influence of environmental conditions, such as the interaction between various oxidizing species in the air and C$_2$N, holds both fundamental and technological significance.

In recent years, numerous research groups have extensively explored the catalytic activity of metal-doped C$_2$N in various gas reduction reactions, both experimentally and theoretically~\cite{he2016iron,cui2018c,ma20163d,mushtaq2021magnetic}. However, a comprehensive understanding of the reaction mechanism between pristine C$_2$N and O$_2$/O$_3$ remains less investigated. In this study, we investigate the interaction between a C$_2$N monolayer and O$_2$/O$_3$ using a state-of-the-art theoretical approach that integrates density functional theory with the climbing image nudged elastic band (CINEB) method~\cite{henkelman2000improved}. Our primary focus is on understanding the impacts of O$_2$ and O$_3$ interaction on the geometrical and electronic structures of C$_2$N by examining chemical bonding, dissociation energy, and density of states~\cite{patra2021ozonation,rawat2024first}.

\section{\label{sec:level2}Computational Methods}
First-principles calculations based on density functional theory (DFT) were performed using the Vienna Ab initio Simulation Package (VASP)~\cite{kresse1996efficient,kresse1996efficiency} with the projector augmented-wave (PAW) method~\cite{blochl1994projector}. The exchange-correlation functional was considered using the Perdew-Burke-Ernzerhof (PBE)~\cite{perdew1996generalized} parametrization-based generalized gradient approximation (GGA) level. The DFT-D2 van der Waals corrections developed by Grimme~\cite{grimme2006semiempirical} were implemented to accurately describe the 2D layer system for the long-range interaction. A vacuum layer of 20 \AA ~was selected along the `Z' direction from the surface to avoid interaction between the periodic layers. The energy cut-off for the plane wave basis was set at 520 eV. Brillouin zone sampling was performed using a $\Gamma$-centred Monkhorst-Pack ($7 \times 7 \times 1$) \textbf{k}-mesh~\cite{monkhorst1976special} for the structural relaxation and ($9 \times 9 \times 1$) \textbf{k}-mesh for the electronic structure calculation of the pristine and adsorbed C$_2$N monolayer. However, a denser ($18 \times 18 \times 1$) \textbf{k}-mesh was considered for the calculations of the projected density of states using the tetrahedron method. The initial separation between the C$_2$N monolayer and the O$_2$/O$_3$ molecule was set at 4 \AA. The structural relaxation was performed for the pristine and adsorbed monolayers until the total energy and the Helman-Feynman forces acting on each atom were less than 10$^{-6}$ eV and 0.01 eV/\AA~ respectively. The O$_2$ interaction was studied using a single C$_2$N monolayer with twelve carbon and six oxygen atoms. In contrast, a larger ($2 \times 2 \times 1$) supercell was considered for the ozonation process. The ionic relaxation was performed by placing the O$_2$/O$_3$ molecules at different locations on the C$_2$N monolayer with both horizontal and vertical orientation. The structure with minimum energy was considered for further calculations. The CINEB method~\cite{henkelman2000improved} was adopted to study molecular interactions with the monolayer surface, describing the minimum energy path between the initial and final configuration. Atomic charges at the adsorbed molecules were calculated using Bader’s atoms-in-molecules approach~\cite{bader1990atoms}. The molecule-C$_2$N interaction binding energies ($E_B$) have been computed as:
\begin{equation}
    E_B= E(molecule+C_2N)-E(molecule)-E(C_2N)
\end{equation}
where $E(molecule+C_2N)$, $E(molecule)$ and $E(C_2N)$ are the total energies of the molecule adsorbed C$_2$N configuration, the interacting molecule, and the pristine C$_2$N monolayer, respectively. The binding energies for the physisorbed configurations are denoted as physisorbed energy ($E_{\text{phys}}$). The difference in energies between the non-interacting configurations and the dissociated configurations is designated as dissociation energy ($E_{\text{diss}}$).

\section{\label{sec:level3}RESULTS AND DISCUSSION:}
\subsection{Structural properties}
Structural relaxations were conducted for both C$_2$N monolayers and O$_2$ and O$_3$ molecules using the PBE-DFT level of theory. The optimized lattice constant for the pristine C$_2$N monolayer was determined to be 8.33 Å, which aligns with prior experimental \cite{mahmood2015nitrogenated} and theoretical investigations \cite{varunaa2019potential, ashwin2017tailoring, Das2024}. The calculated bond lengths for C--C and C--N bonds are approximately 1.43 (1.47) Å and 1.34 Å, respectively. Meanwhile, the optimized bond lengths for O$_2$ and O$_3$ molecules are 1.23 Å and 1.28 Å, respectively, and the bond angle of the O$_3$ molecule is 118.23\textdegree, consistent with previous findings~\cite{moffitt1951electronic, mainali2011first, kalemos2008electronic, patra2021ozonation}.

\begin{figure}[tbp!]
\includegraphics[width=1.0\linewidth]{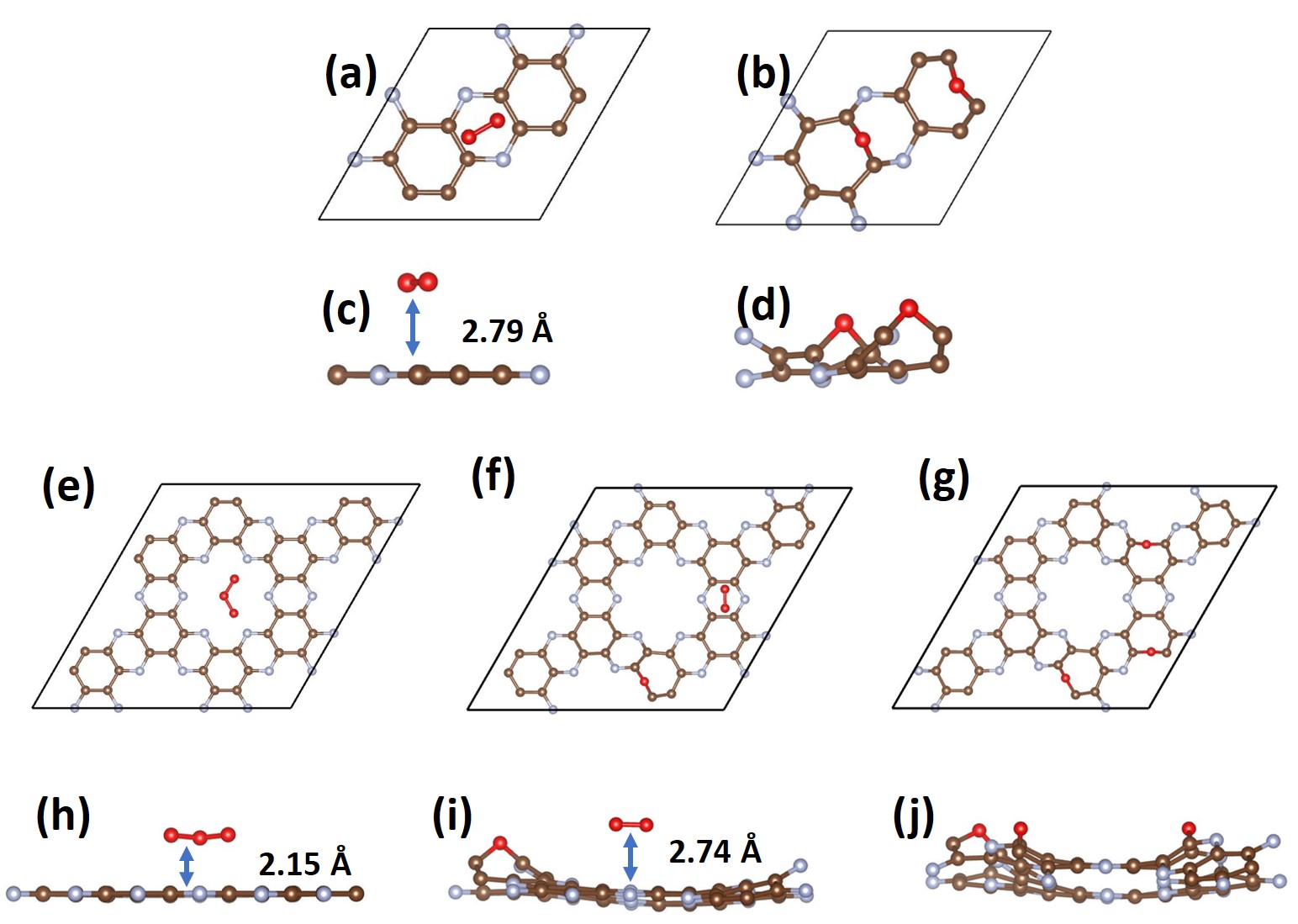}
\caption{\textbf{Structural configurations of O$_2$/O$_3$ interacting with C$_2$N}: Top view of the relaxed configuration of (a) O$_2$ molecule (b) atomic oxygen on C$_2$N monolayer surface. The bottom panel indicates the side view of the corresponding structure.}
         \label{fig:relax}
\end{figure}

The initial separation between the C$_2$N monolayer and the molecules is set to 4 Å at various possible positions across the monolayer (refer to Table S1 for details in the Supporting Information (SI)). Initially, these configurations can be considered as non-interacting, given the small binding energies observed, i.e. 0.03 eV and 0.07 eV for O$_2$ and O$_3$, respectively. Subsequently, these configurations are allowed to fully relax to attain the physisorbed configurations. The total energy values indicate that O$_2$ and O$_3$ molecules oriented horizontally tend to remain positioned above the pyrazine ring (see Fig.~\ref{fig:relax}(a, c)) and the hollow region (see Fig.~\ref{fig:relax}(e, h)) of the monolayer, respectively. These lowest energy configurations are then selected for further investigations.  The optimized distances for the lowest energy configurations are determined to be 2.79 Å (Fig.~\ref{fig:relax}(a, c)) and 2.15 Å (Fig.~\ref{fig:relax}(e, h)) for the O$_2$ and O$_3$ adsorbed systems, respectively.

Figures~\ref{fig:relax}(b) and (d) illustrate the O$_2$ dissociated systems, where the individual O atoms are bonded to carbon atoms on the surfaces, forming epoxy (C-O-C) groups, resembling the configuration observed in graphene and C$_3$N monolayer~\cite{patra2021ozonation,zhao2023oxygen}. Unlike O$_2$, the dissociation of O$_3$ involves two steps. Initially, O$_3$ decomposes into a molecular O$_2$ and an atomic O. The O atom then bonds to C-C bonds, creating a C-O-C group, while the O$_2$ molecule remains floating over the surface at a distance of 2.74 Å (see Fig.~\ref{fig:relax}(f,i)), resembling the O$_2$-pristine C$_2$N physisorbed configuration. In the final step, the O$_2$ molecule further dissociates into two O atoms, with each attached to C-C bonds, resulting in the formation of two additional C-O-C groups as observed in graphene~\cite{patra2021ozonation} (see Fig.~\ref{fig:relax}(g,j)). The creation of epoxy groups involves charge transfers from the monolayer to the atoms, as discussed in the following sections. The detailed information on the geometrical optimization for the adsorption of O$_2$/O$_3$ on C$_2$N monolayer are given in Fig. S1-S6 in the SI.

\begin{figure}[!tbp!]
    \includegraphics[width=0.6\linewidth]{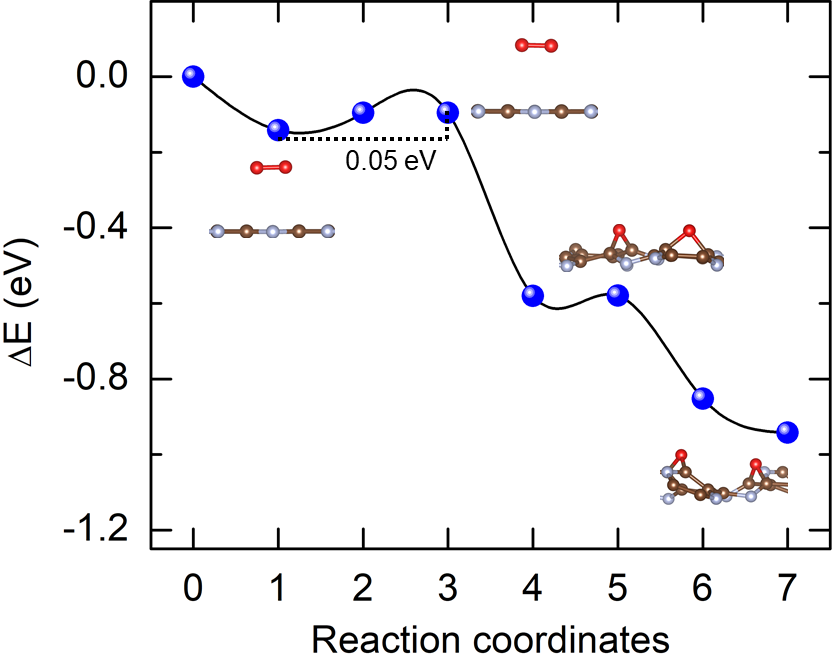}
         \caption{\textbf{Reaction pathway for O$_2$ with C$_2$N}: Adsorption and dissociation process of O$_2$ on the C$_2$N monolayer surface, including structural configurations at intermediate stages. The colour scheme for atoms corresponds to that utilized in Fig.~\ref{fig:relax}.}
         \label{fig:neb-O2}
\end{figure} 

\subsection{Oxygen reactivity}
\begin{figure*}[t!]
\includegraphics[width=0.8\linewidth]{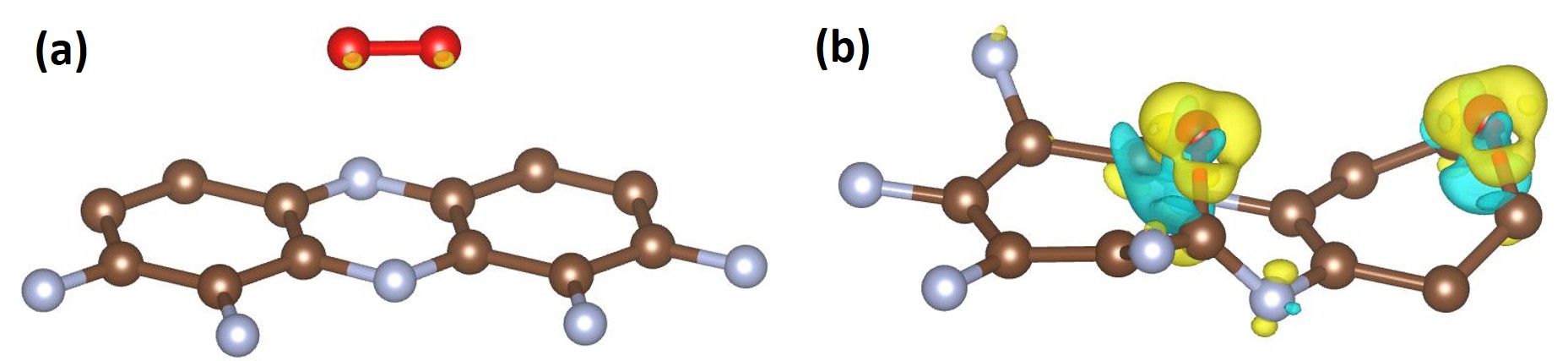}
\caption{\textbf{Charge density difference of O$_2$ on C$_2$N}: (a) physisorbed (O$_2$) and (b) dissociated (O + O) configurations. The iso-surface value was set at 0.03 e/Å$^3$. The yellow and cyan colours represent charge accumulation and depletion, respectively. The colour scheme for atoms corresponds to that utilized in Fig.~\ref{fig:relax}.}
         \label{fig:chg_o2}
\end{figure*}

Figure~\ref{fig:neb-O2} illustrates the transition path depicting the chemical reactivity of O$_2$ on the pristine C$_2$N monolayer. Initially, we positioned the O$_2$ molecule on the top of a 1 × 1 × 1 unit cell of the C$_2$N monolayer, maintaining a vertical distance of 4 Å from the surface. Given the negligibly small calculated binding energy, this configuration serves as a non-interacting state, with energy values for subsequent configurations calculated relative to this state. Following complete ionic relaxation, we observed the distance between the C$_2$N surface and the O$_2$ molecule is reduced to 2.79 Å (see Fig.~\ref{fig:relax}(c)). Additionally, the O-O bond length was found to be 1.24 Å, consistent with the bond length of a free O$_2$ molecule in the triplet state (see Table S2 in the SI). Calculation of the binding energy yielded -0.16 eV, indicating weak physisorption of O$_2$ on pristine C$_2$N surfaces. The molecular adsorption process is further analyzed using Bader charge and charge density difference plot analyses. Our findings reveal that physisorption involves a minor charge transfer ( $\sim 0.12 e$) from the C$_2$N monolayer to the O$_2$ molecule. This is supported by the negligible charge accumulation observed at the O$_2$ site in the charge density difference plot (see Fig.~\ref{fig:chg_o2}(a)). Chakrabarty et al.~\cite{chakrabarty2017electron} also reported similar weak physisorption of O$_2$ on the C$_2$N monolayer. 

Following physisorption, O$_2$ encounters a small energy barrier of 0.05 eV as it proceeds through several intermediate steps, ultimately dissociating the O$_2$ molecule into two O atoms. These dissociated O atoms subsequently attach to the C-C bonds, forming epoxide (C-O-C) groups on the surface, with a C-O bond length of $\sim 1.4$ Å as depicted in Fig.~\ref{fig:relax}(b, d) (see Table 1 for details). Bader’s charge analysis indicates a charge transfer of approximately $1e$ from C$_2$N to the chemisorbed oxygen atoms. This is further supported by the significant accumulation of charges at the O sites, as observed in the charge difference plots provided in Fig.~\ref{fig:chg_o2}(b). The dissociation of O$_2$ molecules into two O atoms by accepting charges from the C$_2$N monolayer can be attributed to the antibonding nature of its lowest unoccupied molecular orbital (LUMO)~\cite{patra2022surface}.

\subsection{Ozone reactivity}

\begin{figure*}[t!]
\includegraphics[width=1\linewidth]{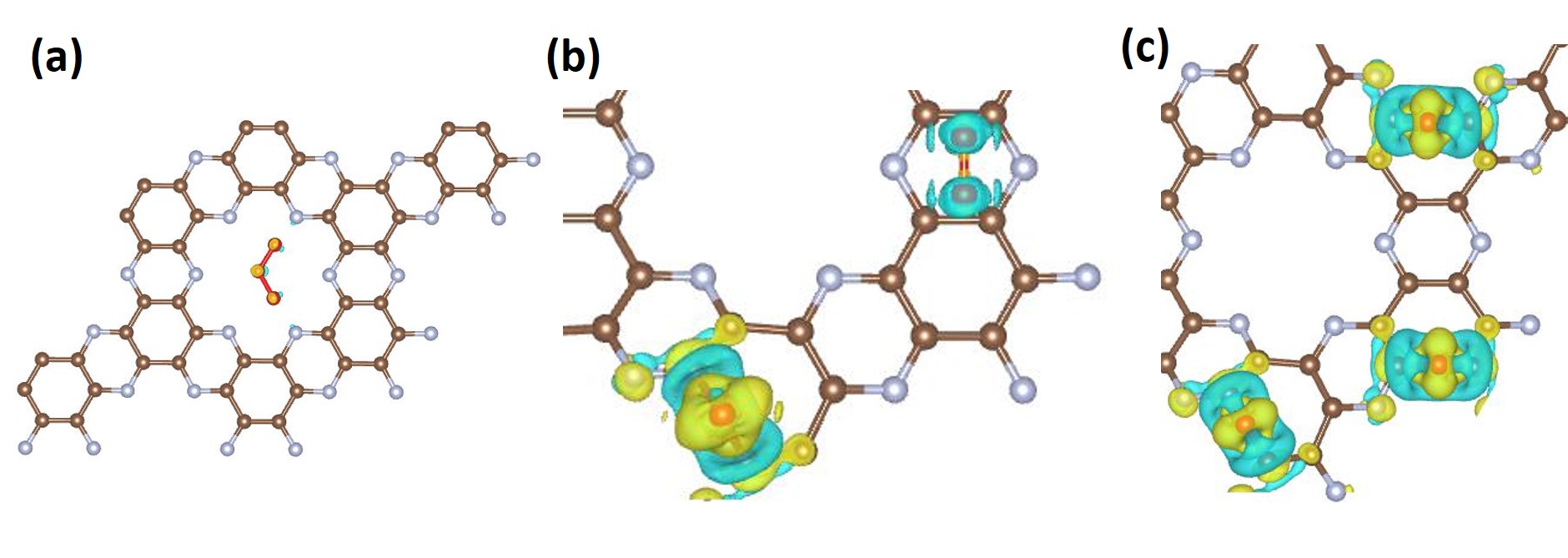}
\caption{\textbf{Charge density difference of O$_3$ on C$_2$N}: (a) physisorbed (O$_3$), (b) intermediate (O$_2$ + O), and (c) dissociated (O + O + O) configurations. The iso-surface value for the O$_3$ calculation was set at 0.09 e/Å$^3$. The yellow and cyan colours represent charge accumulation and depletion, respectively. The colour scheme for atoms corresponds to that utilized in Fig.~\ref{fig:relax}.}
         \label{fig:chg_o3}
\end{figure*}

Similar to O$_2$, we investigated the reactivity of O$_3$ by placing it 4 Å above the C$_2$N monolayer. Following full relaxation, the O$_3$ molecule relaxed to a distance of 2.15 Å above the hollow region, as illustrated in Figure~\ref{fig:relax}(e, h). The relaxed configuration exhibited a weak binding energy of -0.31 eV, indicating the presence of physical adsorption. Bader charge analysis suggests that physisorption is facilitated by a minimal charge transfer of approximately $\sim 0.08 e$, which is confirmed by the negligible accumulation of charges on the O$_3$ molecule (Fig.~\ref{fig:chg_o3}(a)). The dissociation of ozone molecules occurs in two steps: O$_3 \rightarrow$ O$_2$ + O followed by O + O + O. Interestingly, this process is similar to the ozone dissociation observed in graphene monolayer~\cite{patra2021ozonation}.

\begin{figure}[!tbp!]
\includegraphics[width=1\linewidth]{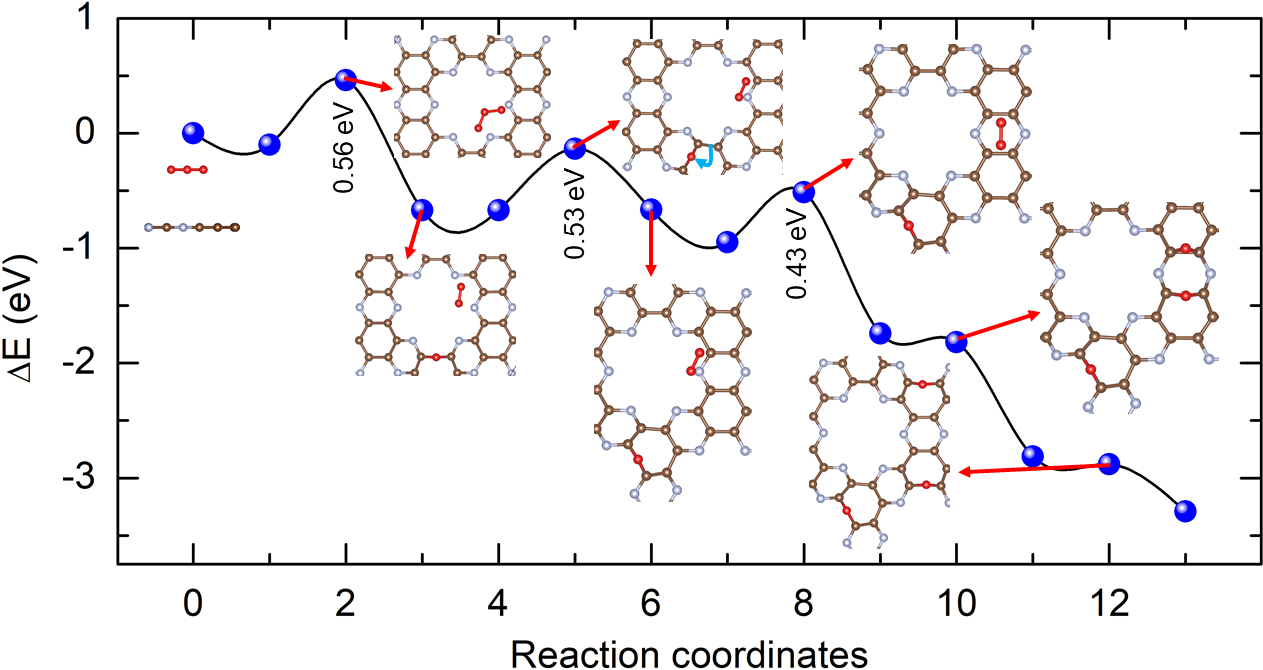}
\caption{\textbf{Reaction pathway for O$_3$ with C$_2$N}: Adsorption and dissociation process of O$_3$ on the C$_2$N monolayer surface, including structural configurations at intermediate stages. The cyan arrow indicates the trajectory of the O atom as it transitions between C-C bonds. The colour scheme for atoms corresponds to that utilized in Fig.~\ref{fig:relax}.}
\label{fig:neb-O3}
\end{figure}

It is well-established that O$_3$ possesses longer and substantially weaker bonds compared to O$_2$ due to the coupling of a pair of electrons in orbitals on each of the terminal atoms, representing a weak $\pi$ interaction within the molecule~\cite{takeshita2015insights}. Consequently, when O$_3$ interacts with C$_2$N monolayers, it initiates bond breaking in the molecule, yielding O and O$_2$, before further dissociating into atomic oxygens. The complete transition path for O$_3$ adsorption and dissociation is depicted in Fig.~\ref{fig:neb-O3}. Notably, the O$_3$ molecule must traverse three energy barriers during the dissociation process. Following relaxation on the surface, O$_3$ begins moving towards the border of the hollow area from the centre, resulting in a higher energy metastable state. This state, higher in energy than its previous intermediate state, leads to the first energy barrier of 0.56 eV.

Subsequently, O$_3$ breaks into O + O$_2$, and the O$_2$ molecule remains at a height of 2.74 Å, indicating a physisorbed state. This height, slightly less than that in the O$_2$ physisorbed pristine C$_2$N case (2.79 Å), can be attributed to the attached O atom to the C$_2$N monolayer suggesting a stronger interaction. The Bader charge analysis and the charge density difference plots (Fig.~\ref{fig:chg_o3}(b)) reveal a charge accumulation of approximately $\sim 1e$ at the O site. The dissociated O atom then relaxes from the C-C bond of the hollow site to the C-C bond of the pyrazine ring (indicated by a cyan arrow in Fig.~\ref{fig:neb-O3}) to find a position of minimum energy. This jump leads to the second energy barrier of 0.53 eV. Subsequently, the dissociated O and O$_2$ on the C$_2$N monolayer undergo complete relaxation, marking the completion of the first step of the dissociation process.

\begin{table}[!hb]
  \centering
  \caption{Calculated physisorption energy (E\textsubscript{phys}), dissociation energy (E\textsubscript{diss}), C-O bond lengths (R\textsubscript{C-O}), Bader charges at O sites (Q\textsubscript{0}/O), and energy barrier values (E\textsubscript{barrier}) of the initial, intermediate, and final configurations describing O$_2$/O$_3$ interaction with the C$_2$N monolayer}
    \begin{tabular}{ccccccc}
    \toprule
    & \multicolumn{2}{c}{O$_2$} & \multicolumn{3}{c}{O$_3$} \\
    \cmidrule(lr){2-3} \cmidrule(lr){4-6}
    \textbf{Process} & \textbf{Physisorption} & \textbf{Dissociation} & \textbf{Physisorption} & \textbf{intermediate }&\textbf{Dissociation} \\
                   & (O$_2$) & (O + O) & (O$_3$) & (O + O$_2$) & (O + O + O)\\
    \midrule
    E\textsubscript{phys}, eV & -0.16& --- & -0.31& ---&--- \\
    E\textsubscript{diss}, eV & ---& -0.94& ---&-0.95& -3.29\\
    R\textsubscript{C-O}, Å & 2.79 & 1.39 & 2.15&1.39, 2.74& 1.38\\
    Q\textsubscript{0}/O atom, $e$ & 0.05 & 1.04 & 0.03 &1.08, 0.05& 1.07 \\
    E\textsubscript{barrier}, eV & ---& 0.05& ---&0.56, 0.53, 0.43&---  \\
    \bottomrule
    \end{tabular}%
  \label{tab:energy}%
\end{table}%

In the next step, the O$_2$ molecule moves into the pyrazine ring by crossing the third energy barrier of 0.43 eV. Interestingly, unlike in the pristine case, the O$_2$ molecule then breaks into two oxygen atoms instantaneously without encountering any energy barrier. This observation suggests that the oxidized C$_2$N monolayer is more reactive to O$_2$ exposure. Following this, the dissociated O atoms attach to the two C-C bonds of the pyrazine ring. However, the O atoms undergo relaxation and repel each other away, driven by the Coulomb interaction between their p orbitals, eventually attaching to the benzene rings (Fig.~\ref{fig:neb-O3}). A similar repulsion phenomenon is observed among chemisorbed oxygen atoms on the phosphorene surface~\cite{ziletti2015oxygen}. The average C-O bond lengths are approximately $\sim 1.4$ Å for the epoxide (C-O-C) groups in the final relaxed state. Table~\ref{tab:energy} provides a summary of the physisorption (E\textsubscript{phys}), dissociation energy (E\textsubscript{diss}), C-O bond lengths (R\textsubscript{C-O}), Bader charges at O sites (Q\textsubscript{0}/O), and energy barrier values (E\textsubscript{barrier}) associated with the reactivity of O$_2$ and O$_3$ with the C$_2$N monolayer

\subsection{Electronic Structure Analysis}
\begin{figure}[tbp!]
\includegraphics[width=1\linewidth]{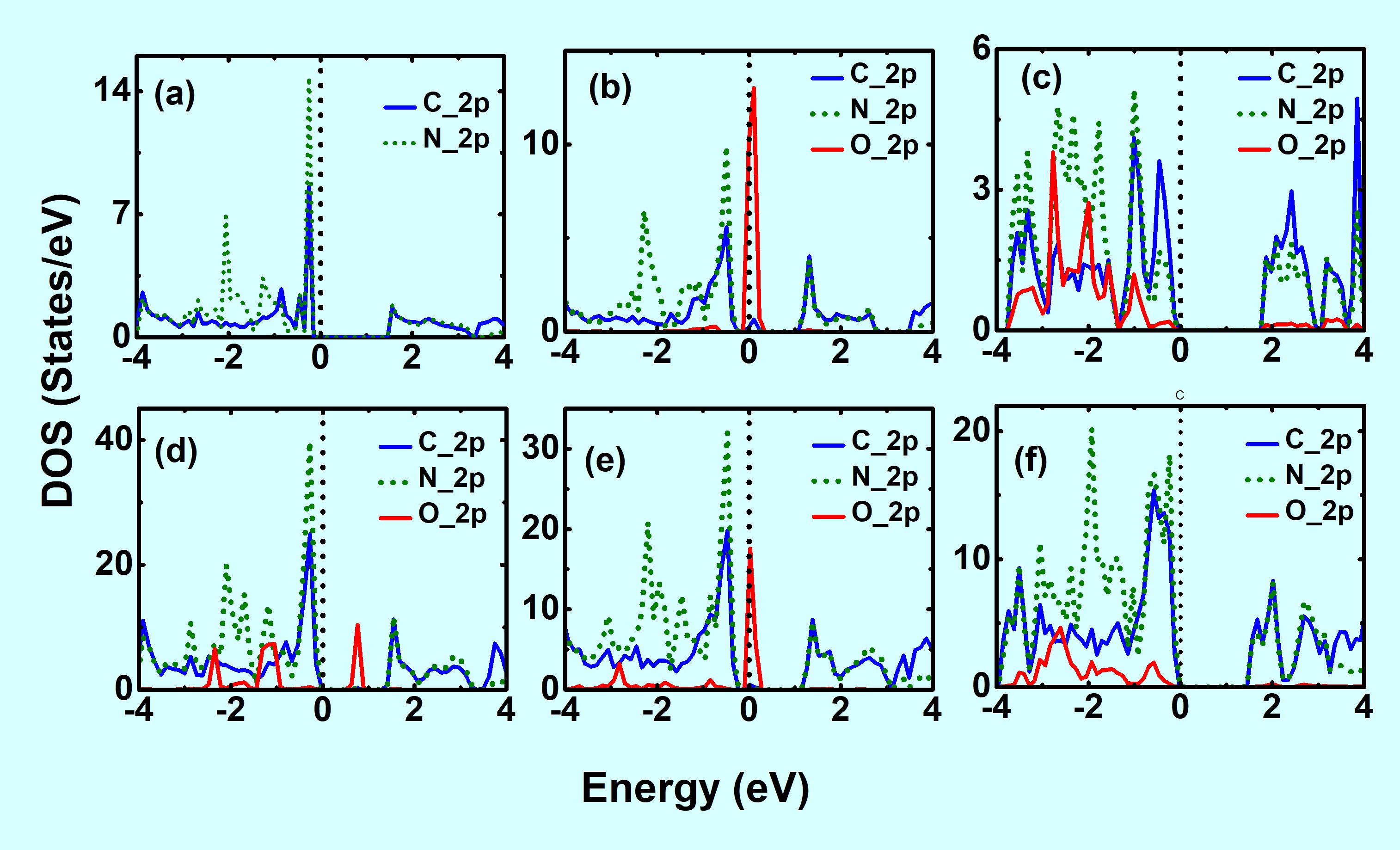}
\caption{\textbf{Projected density of states for O$_2$/O$_3$ interaction with C$_2$N}: Top panel: (a) pristine C$_2$N, (b) physisorbed (O$_2$), and (c) dissociated (O + O) oxygen configurations. Bottom panel: (a) physisorbed (O$_3$) (b) intermediate (O$_2$ + O), and (c) dissociated ( O + O + O) ozone configurations.}
\label{fig:dos}
\end{figure}

 Using the PBE-DFT level of theory, a direct band gap of 1.66 eV is estimated for the pristine C$_2$N monolayer (refer to Fig.~\ref{fig:dos}(a)), consistent with previous reports \cite{mahmood2015nitrogenated,ashwin2017tailoring}. The projected density of states (PDOS) for the pristine C$_2$N monolayer Fig.~\ref{fig:dos}(a) show that the C and N `p' orbitals majorly contribute to the valence bands around the Fermi level (E\textsubscript{F}). Physisorption of O$_2$ on the C$_2$N monolayer does not significantly alter the band dispersions; instead, it forms localized molecular states around the E\textsubscript{F} (Fig.~\ref{fig:dos}(b)). In contrast, the dissociated configuration exhibits hybridized oxygen-C$_2$N bands dispersed deep inside the valence band region, nearly 2 eV below the E\textsubscript{F}. This dissociated configuration shows an increased direct band gap of 2.09 eV, which is ascribed to robust C-O hybridization within the epoxy (C-O-C) group on the surface (Fig.~\ref{fig:dos}(c)).

In the O$_3$ physisorbed configuration, localized states are observed, indicating minimal interaction between the molecule and the monolayer (Fig.~\ref{fig:dos}(d)). Upon dissociation of the O$_3$ molecule into O$_2$ + O, the resulting separated O atoms hybridize with C atoms within the C$_2$N monolayer, forming hybridized bands situated below 3 eV from the  E\textsubscript{F}. In contrast, the weakly absorbed O$_2$ molecule yields sharp peaks at the E\textsubscript{F} (Fig.~\ref{fig:dos}(e)). Following the final step, dissociated O atoms establish stronger bonds with C atoms on the C$_2$N surface, generating hybridized bands extending deep below the E\textsubscript{F} (Fig.~\ref{fig:dos}(f)). This dissociated configuration demonstrates an increased band gap (1.76 eV) as compared to the pristine case (1.66 eV). These observations are further evidenced from the electronic band structure calculations given in Fig. S7, in the SI. Our findings indicate that the adsorption of O$_2$ and O$_3$ can significantly modify the electronic structure of the C$_2$N monolayer, primarily due to the formation of epoxy groups on the C$_2$N surface.

\section{\label{sec:level4} CONCLUSION:}
We conducted a systematic investigation into the interactions between O$_2$ and O$_3$ molecules and C$_2$N monolayers, employing DFT along with the CINEB approach. Our findings reveal that C$_2$N exhibits stability under exposure to O$_2$ and O$_3$, with energy barriers of 0.05 eV and 0.56 eV, respectively. Upon interaction with O$_2$, dissociation occurs, leading to the formation of epoxy groups (C-O-C) on the surface through the bonding of two O atoms. Conversely, ozonation proceeds in two stages: firstly, O$_3$ dissociates into an atomic O and molecular O$_2$, followed by further dissociation of O$_2$ into two O atoms, resulting in the creation of three epoxy groups on the C$_2$N surface. Analyses based on charge density difference and Bader charge suggest that the dissociation process involves significant charge transfers from the surface to the molecules. Examination of the electronic structure reveals higher bandgaps in the dissociated configurations, attributed to C-O hybridizations. Overall, our study indicates that C$_2$N monolayers can undergo oxidation by O$_2$/O$_3$, overcoming energy barriers, and resulting in enhanced bandgaps. Overall, the electronic properties of C$_2$N can be tailored by adjusting the level of O$_2$/O$_3$ exposure, offering potential avenues for the design of electronic devices based on C$_2$N and facilitating advancements in various technological domains.

\begin{acknowledgments}
The authors would like to thank the National Institute of Science Education and Research (NISER), Department of Atomic Energy, Government of India, for funding the research work through project number RIN-4001. The authors acknowledge the high-performance computing facility at NISER.
\end{acknowledgments}

\nocite{*}
\bibliography{ref}

\end{document}